\documentclass[twocolumn,prl, nofootinbib, superscriptaddress, preprintnumbers]{revtex4}

\usepackage{graphicx}
\usepackage{epsfig}
\usepackage{amsmath, amssymb}
\usepackage{verbatim}
\usepackage{bm}

\newcommand{\gammabar}{\ensuremath\gamma\kern-0.53em-}

\preprint{MIT-CTP 4341}
\begin{document}

\title{A continuous transition between fractional quantum Hall and superfluid states}
\author{Maissam Barkeshli}
\affiliation{Department of Physics, Stanford University, Stanford, CA 94305 }
\author{John McGreevy}
\affiliation{Department of Physics, Massachusetts Institute of Technology, Cambridge, MA 02139 }

\begin{abstract}
We develop a theory of a direct, continuous quantum phase transition between a bosonic Laughlin fractional quantum Hall (FQH) 
state and a superfluid, generalizing the Mott insulator to superfluid phase diagram of bosons
to allow for the breaking of time-reversal symmetry. 
The direct transition can be protected
by a spatial symmetry, and the critical theory is a pair of Dirac fermion fields coupled to an emergent Chern-Simons gauge field. 
The transition may be achieved in optical traps of ultracold atoms by starting with a $\nu = 1/2$ bosonic Laughlin state 
and tuning an appropriate periodic potential to change the topology of the composite fermion band structure. 
\end{abstract}

\maketitle

\it Introduction \rm -- One of the most celebrated examples of a continuous quantum phase transition is between
a Mott insulator (MI) and a superfluid (SF) of bosons \cite{FW8946,GM0239}.
Over the last two decades, this transition has been successfully characterized, both theoretically and experimentally. 
In addition to the Mott insulator and the superfluid, it is expected that a fractional quantum Hall (FQH)
state can be realized in strongly interacting bosonic systems, such as in optical traps of ultracold 
atomic gases \cite{ucaProposals}.
This raises a fundamental question of whether it is also 
possible to transition continuously between FQH 
states and Mott insulators or superfluids. While theories of continuous transitions between FQH states 
and Mott insulators have been developed \cite{ZH8982,WW9301,CF9349,KL9223}, 
it has not been addressed whether the FQH state can 
directly and continuously transition to a superfluid as the kinetic energy of the bosons is increased relative to their interaction energy.

\begin{figure}
\centerline{
\includegraphics[width=3.3in]{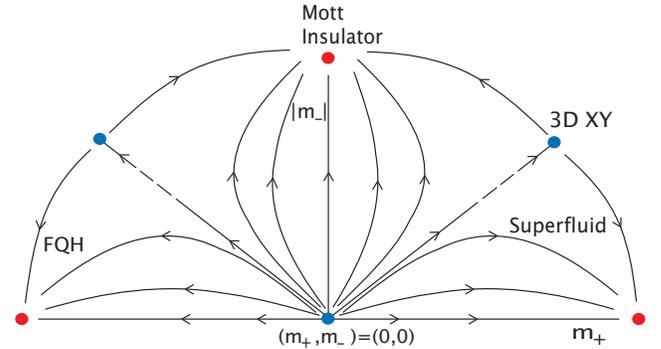}
}
\caption{Proposed phase diagram and renormalization-group flows including the Mott insulator, superfluid, and 
$\nu = 1/2$ Laughlin FQH state, for fixed average particle number. We have defined
$m_\pm \equiv m_1 \pm m_2 $ (see eq. \ref{criticalTh1}); $m_-$ is a symmetry-breaking field, 
so the direct transition between the FQH state and the SF can occur if the symmetry is preserved. 
The red points on the horizontal and vertical axes indicate the three stable phases, while the blue points at the origin 
and the diagonals indicate the unstable critical fixed points. 
\label{phaseDiag}
}
\end{figure}

In this paper, we develop a theory of such a continuous transition, between a $\nu =1/2$ bosonic Laughlin state and 
a superfluid, thereby providing a more general picture of the boson phase diagram (Fig. \ref{phaseDiag}). Since the superfluid is described by an order parameter
while the FQH state is a topological phase without a local order parameter, such a transition is conceptually quite exotic \cite{lroTolre}. 
Realizing it in the lab would be an experimental example of a continuous quantum transition 
in a clean system (unlike QH plateau transitions) which lies outside the Ginzburg-Landau paradigm. 
Here, we will specialize to the case with fixed average particle number.
We find that generically, in the absence of any additional symmetries besides particle number conservation, continuous transitions 
occur between the FQH state and Mott insulator or the Mott insulator and the superfluid. However in the presence of certain 
spatial symmetries, there may be a direct, continuous transition between the FQH state and the superfluid. 

A simple way to understand the basic idea is through the composite fermion \cite{J8999} framework. The $\nu  =1/2$ Laughlin state
can be understood in terms of composite fermions attached to one flux quantum 
each, such that the mean-field state of 
composite fermions is a $\nu  = 1$ integer quantum Hall (IQH) state. 
An externally-applied periodic 
potential can change the band structure of the composite fermions 
such that they occupy bands with a total Chern 
number $C$. When $C = 1$, the state is still the $\nu = 1/2$ FQH state. However, when $C = 0$, the resulting state is 
a Mott insulator, and, as we explain below, when $C = -1$, the resulting state is a superfluid. Thus the transitions 
between these states can be understood as Chern number-changing transitions of the composite fermions. The critical 
theories for such transitions consist of gapless Dirac fermions coupled to a Chern-Simons (CS) gauge field. 


\it Effective field theory constructions \rm -- In order to develop our theory, we need to provide a field theoretic
description that can naturally interpolate between the states of interest. To do this, we will use the parton/projective
construction \cite{parton}. 
For the Laughlin FQH state, the Mott insulator, and the superfluid, the parton construction is essentially
equivalent to the composite fermion construction, although the former is preferable because it can describe a wider 
class of FQH states \cite{partonExotic} 
and can be formulated even in the 
absence of a background external magnetic field \cite{partonZeroField}.   
In this paper we will consider the situation
where the bosons feel an external magnetic field, because it is more directly relevant to ultracold atom proposals, though
the theory can be generalized to cases without an external magnetic field. 
We write the boson operator $b(r)$ as
\begin{align}
b(r) = f_1(r) f_2(r),
\end{align}
where $f_1$ and $f_2$ are charge $1/2$ fermions. This construction introduces an $SU(2)$ gauge
symmetry \cite{wen04}. Since the $f_i$ carry charge $1/2$, they effectively see half as much magnetic field; thus
for bosons at $\nu = 1/2$, the density of $f_i$ is such that their effective filling fraction is $\nu_{f_i} = 1$. To describe the
$\nu = 1/2$ Laughlin state, we assume a mean-field ansatz that breaks the $SU(2)$ gauge symmetry to $U(1)$
and where $f_i$ form $\nu_{f_i} = 1$ IQH states. Letting $a$ denote the emergent $U(1)$ gauge field and $A$
the background external gauge field, integrating out $f_1$ and 
relabelling
$a \rightarrow a + \frac{1}{2} A$ gives
\begin{align}
\label{cfaction}
\mathcal{L} = f_2^\dagger i D_0 f_2 - \frac{1}{2m_{eff}} f_2^\dagger D^2 f_2 + \frac{1}{4\pi} \epsilon^{\mu \nu \lambda} a_\mu \partial_\nu a_\lambda 
+ \delta \mathcal{L},
\end{align}
where the covariant derivative is $D_\mu = \partial_\mu - i a_\mu - i A_\mu$, and 
$\delta \mathcal{L}$ includes additional interactions, external potentials, 
\it etc\rm. This is the same theory obtained by the flux attachment and flux smearing mean-field approximation in the
composite fermion theory, where $f_2$ is the composite fermion. At energies well below the gap of the 
$f_1$ state, a hole of $f_1$ can be created by inserting $2\pi$ flux; thus, for energies below the gap of the $f_1$ 
state, the boson $b$ can be represented by the operator
\begin{align}
b = \hat{M} f_2,
\end{align}
where $\hat{M}$ is an instanton operator that creates $2\pi$ flux of $a$. 
Integrating out $f_2$, which is assumed to form a $\nu_{f_2}=1$ IQH state, and relabelling
$a \rightarrow a - \frac{1}{2} A$ leads to the following effective 
action, to lowest order in the gauge fields and their derivatives:
\begin{align}
\mathcal{L} = \frac{2}{4\pi} \epsilon^{\mu \nu \lambda} a_\mu \partial_\nu a_\lambda +\frac{1}{2} \frac{1}{4\pi} 
\epsilon^{\mu \nu \lambda} A_\mu \partial_\nu A_\lambda .
\end{align}
This gives the correct Hall conductance and reproduces the correct topological degeneracies of the $\nu  =1/2$ Laughlin 
state \cite{wen04}.

Now suppose that $\delta\mathcal{L}$ is 
chosen in such a way that the lowest band for $f_2$
has a general Chern number, $C$. Integrating out the fermions results in the following effective theory, to lowest order:
\begin{align}
\label{genAction}
\mathcal{L} = \epsilon^{\mu \nu \lambda}  \left[ \frac{C+1}{4\pi} a_\mu \partial_\nu a_\lambda + 
\frac{C}{4\pi} A_\mu \partial_\nu A_\lambda + \frac{C}{2\pi} A_\mu \partial_\nu a_\lambda \right].
\end{align}
When $C = 0$, (\ref{genAction}) is simply $\mathcal{L} = \frac{1}{4\pi} \epsilon^{\mu \nu \lambda} a_\mu \partial_\nu a_\lambda $, 
which describes a gapped state with a unique ground state on all closed manifolds. The gapped $f_2$
excitations are attached to a unit of flux, so they are bosonic excitations. A careful analysis 
following \cite{parton} reveals there are no gapless protected edge states. Such a gapped state
with solely bosonic excitations and unique ground state degeneracies is a Mott insulator. This result
can also be cast within the composite boson language \cite{ZH8982}, where the original boson is considered to be
a composite boson $\phi$ attached to two units of flux. Performing the flux smearing approximation gives
composite bosons in no net magnetic field. The $\langle \phi \rangle \neq 0$ and $\langle \phi \rangle = 0$ 
states correspond to the FQH state and Mott insulator, respectively. This is just the bosonized
description of the $C = 1$ and $C = 0$ composite fermion description of these states. 

Since $a$ is a dynamical gauge field, to describe a gapped state,
the gauge fluctuations must be gapped and, to describe a fractionalized state, the gauge theory
must be at a deconfined fixed point. Since CS gauge theories are gapped \cite{DJ8272} and 
represent deconfined quantum field theories \cite{W8951,FS9176}, the above construction can be used to represent FQH states. 
However, when $C = -1$, from (\ref{genAction}) we see that there is no CS term for $a$. 
Restoring
the Maxwell terms to (\ref{genAction}), the effective action is perturbatively, to lowest order, given by
\begin{align}
\label{sfAction}
\mathcal{L} &= \frac{1}{2\pi} \epsilon^{\mu \nu \lambda} A_\mu \partial_\nu a_\lambda + \frac{1}{g_1^2} f^2+ 
\frac{1}{g_2^2} F^2 + \frac{1}{g_3^2} f F,
\end{align}
where the Maxwell term is $f^2 \equiv f_{\mu \nu} f^{\mu \nu}$, and similarly for the last two terms, and we have
assumed Lorentz invariance for simplicity. 
Since there is no CS term $\epsilon^{\mu \nu \lambda} a_\mu \partial_\nu a_\lambda$, 
we must reconsider whether the gauge fluctuations are gapped. 
Without the CS term, in 2+1 dimensional compact $U(1)$ gauge theory, instantons 
proliferate and condense at low energies, 
yielding a contribution $e^{-S_0} \hat{M} + H.c.$ to the effective action \cite{P7729}.
This induces a gap for $a$. However this term cannot be added to (\ref{sfAction}). From the mutual CS 
term $\epsilon^{\mu \nu \lambda} A_\mu \partial_\nu a_\lambda$, we see that flux of $a$ carries 
electric charge. $\hat{M}$, which instantly adds $2\pi$ flux, instantly causes a local depletion 
of the charge density; to satisfy charge conservation, it must create 
a current $j \sim \delta(t)$, which costs an infinite action. Thus instantons alone are suppressed at energies 
below the gap of the fermion states \cite{RV0821, GS0804}. Since $\hat{M}$ creates a hole in the parton IQH states, 
the only possible instanton term that might be added to the effective action at low energies, below
the fermion gap, is of the form $\hat{M} f_1^\dagger f_2^\dagger + H.c.$. The fermion operators fill in
the hole created by the flux insertion, thus keeping the charge density uniform. Such a term is gauge-invariant
if, under a gauge transformation $f_i \rightarrow e^{i\gamma/2} f_i$, $A \rightarrow A - \partial \gamma$,
$\hat{M} \rightarrow e^{i \gamma} \hat{M}$. Such a term does not gap out the gauge field, and
leads to spontaneous symmetry breaking of the fermion number conservation \cite{AH8213}.
Proliferation of these allowed instantons may be viewed as the mechanism within the gauge theory by which
the fermion number conservation is spontaneously broken \cite{AH8213}. 

From the action (\ref{sfAction}), we see that magnetic fluctuations of $a$ are charged under the external
gauge field, which implies that they correspond to density fluctuations \cite{RV0821}. Thus $a$ is dual to the superfluid Goldstone mode.
In fact, (\ref{sfAction}) is dual to the standard superfluid action, as can be seen by introducing 
$\xi^\mu \equiv \frac{1}{2\pi} \epsilon^{\mu \nu \lambda} \partial_\nu a_\lambda$ and a Lagrange multiplier 
$\varphi$ to enforce the constraint $\partial_\mu \xi^\mu = 0$, and subsequently
integrating out $\xi_\mu$. This yields $\mathcal{L} \propto (\partial \varphi - A)^2$. Alternatively, integrating out $a$
in (\ref{sfAction}) yields the standard superfluid response $\mathcal{L} \propto A_\mu (\delta_{\mu \nu} - \frac{p_\mu p_\nu}{p^2}) A_\nu$. 
We conclude that when $f_2$ fills bands with $C=-1$, 
the resulting state is a  superfluid.\footnote{While this appears surprising, we note that it is implicit in 
\cite{CF9349}, where it was argued that the 3D XY critical point can be fermionized. However, 
where there is overlap, some of our results differ from those of \cite{CF9349}. Similarly, \cite{RV0821} uses an 
equivalent construction in a different context, for an $XY$ N\'eel state.} 
In the Supplementary Materials, we give a further discussion of how such a construction can describe a compressible state. 

We note that within this effective field theory description, a deformation of the composite fermion 
bandstructure that causes the bands to overlap will result in a compressible non-Fermi 
liquid state, with a composite fermion Fermi surface \cite{HL9312}.

\it Critical theory \rm -- The critical theories between the FQH state, MI, and SF therefore occur when the 
composite fermion $f_2$ bands touch and their net total Chern number changes. The transition between the SF 
and the $\nu = 1/2$ FQH state occurs when the total Chern number of $f_2$ changes from $1$ to $-1$. 
This can happen either at a quadratic band touching or at two Dirac cones; the generic, stable case is two Dirac cones, 
because quadratic band touchings are marginally unstable to repulsive interactions \cite{SY0911}. To describe this,
let $\psi(r)$ be a two-component fermion that describes the two $f_2$ bands that are involved in the transition, so that
at low energies, $f_2(r) \sim c^T(r) \psi(r)$, where $c(r)$ is a two-component scalar function of $r$; \it ie \rm
at low energies $f_2(r)$ is a linear combination of the two bands described by $\psi$. 
Near the transition, at low energies $\psi(r) \sim \sum_{i=1}^2 e^{i K_i r} \psi_i(r)$, where the Dirac points occur at momenta 
$K_i$ and $\psi_i$ are the two-component fermions obtained by linearizing about the Dirac points. The critical theory is:
\begin{align}
\label{criticalTh1}
\mathcal{L} = \frac{1}{4\pi} \epsilon^{\mu \nu \lambda} a_\mu \partial_\nu a_\lambda + \bar{\psi}_i \gamma^\mu D_\mu \psi_i + m_i \bar{\psi}_i \psi_i,
\end{align}
for $i = 1,2$, $\bar{\psi}_i = \psi^\dagger_i \sigma^z$, $\gamma_0 = \sigma_z$, $\gamma_x = \sigma_x$,
$\gamma_y = \sigma_y$, where $\sigma_i$ are the Pauli matrices. 
When both $m_i < 0$, we obtain the superfluid state, when $m_i > 0$, 
we obtain the FQH state, and if $m_i$ have opposite signs, then we have the Mott insulator (see Fig. \ref{phaseDiag}).

Critical points occur when some $m_ i =0$.\footnote{Note that in addition, 
chemical potential terms $\mu_i \psi^\dagger_i \psi_i$ are relevant operators that lead to a composite Fermi 
liquid. Nevertheless, spatial symmetries can impose $\mu_i = \mu$, and if particle number is held fixed,
as in cold atoms settings, the composite Fermi liquid can be avoided and one can tune through these transitions with a single mass parameter.}
In the absence of any symmetries, the generic transition from FQH to SF 
therefore is through the Mott insulator. However, 
certain spatial symmetries may force $m_1 = m_2$ (see below), in which case there is a single tuning
parameter that tunes between the superfluid and the FQH state. 
\def\kk{k}

Integrating out a Dirac fermion with mass $m$ coupled to a gauge field $a$ yields a CS term
$\frac{\text{sgn} (m) }{2}\frac{1}{4\pi} \epsilon^{\mu \nu \lambda} a_\mu \partial_\nu a_\lambda$. Thus, we consider
the following Lagrangian \cite{redlich}:
\begin{align}
\label{criticalThNf}
\mathcal{L}_{N_f, \kk} = \frac{N_f \kk}{4\pi} 
\epsilon^{\mu \nu \lambda} a_\mu \partial_\nu a_\lambda + \sum_{i=1}^{N_f} [\bar{\psi}_i \gamma^\mu D_\mu \psi_i + m \bar{\psi}_i \psi_i].
\end{align}
The MI-SF transition is described by $\mathcal{L}_{1,1/2}$, the FQH-MI transition is described by $\mathcal{L}_{1,3/2}$,
and the FQH-SF transition is described by $\mathcal{L}_{2,1/2}$ (see Fig. \ref{phaseDiag}). This ``fermionization'' of the 
3D XY transition was already conjectured in \cite{CF9349}. A crucial point is that the FQH-MI transition is different from
the MI-SF transition because of the coefficient of the CS term, which affects the critical properties \cite{WW9301,CF9349}.

The critical exponents can be computed through a large $N_f$ expansion, which has already been performed \cite{CF9349}, 
motivated by the case $N_f = 1$. This is a relativistic transition, with dynamic critical exponent $z = 1$. The 
correlation length exponent $\nu$ is defined by $\xi \sim m^{-\nu}$, where $\xi$ is the correlation length 
and $m$ is the tuning parameter. $\nu$ can be determined by the dimension of the mass term. In the large $N_f$ limit, it was found to be
\begin{align}
\nu^{-1} = 1 + \frac{128}{3}\frac{[128 - (\pi/\kk)^2] }{[64 + (\pi/\kk)^2]^2\kk^2}\frac{1}{N_f} + O(1/N_f^2),
\end{align}
although for $N_f = 1$ the leading $1/N_f$ correction was found \cite{CF9349}
to be insufficient for accurately giving the 3D XY value of $\nu^{-1} \sim 1.5$.
For the FQH-SF transition, $N_f = 2$, $\kk = 1/2$, we expect the large $N_f$ expansion
to be more reliable, and we get $\nu^{-1} = 1.705... + O(1/N_f^2)$.

At low energies the boson operator is $b \sim \hat{M} \psi$, so the scaling dimension $\Delta_b$ of $b$ must be found by 
analyzing the dimension of the monopole operator combined with the fermion. If there are $N_f$ Dirac points in the 
Brillouin zone, at momenta $K_i$, for $i = 1, \cdots, N_f$, then $\psi(r) \sim \sum_i e^{i K_i r}\psi_i(r)$. 
So far, the scaling dimension of an operator like $\hat{M} \psi_i$ is known only in the $N_f \rightarrow \infty$ limit. 
In that limit, the scaling dimension of $b$ is $\Delta_b = \Delta_M + \Delta_\psi$. Furthermore, in the large $N_f$ limit,
$\Delta_M$ can be computed from the state-operator correspondence \cite{BK0249}, 
with the result $\Delta_M = N_f (0.265...)$, while $\Delta_\psi = 1 + O(1/N_f)$. 

The order parameter exponent $\beta$ for the superfluid is defined by $\langle b \rangle \sim m^{\beta}$.
Following the arguments in \cite{FW8946}, $\beta$ can be seen to obey a generalized hyperscaling relation: $\beta = \nu \Delta_b$.
The large $\Delta_b$ implies that the onset of superfluidity is quite weak at the FQH-SF transition. 
Additionally, from general hyperscaling arguments, the superfluid susceptiblity scales like $\chi \sim m^{\nu(2\Delta_b-d-z)}$.
If we naively plug in $\Delta_b = 0.265 N_f + 1$, then for $N_f = 2$, $\chi$ is non-divergent at the critical 
point.\footnote{We thank T. Senthil for emphasizing this point.}
\begin{figure}[t]
\centerline{
\includegraphics[width=3.4in]{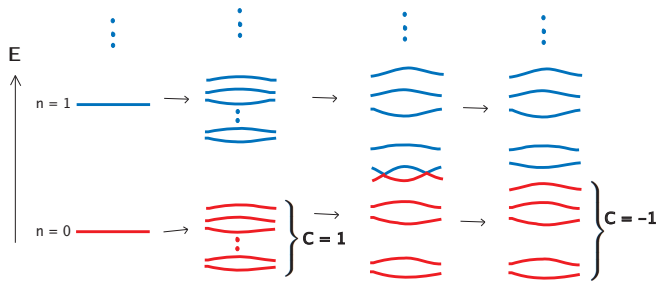}
}
\caption{Evolution of composite fermion bands as a periodic potential is turned on and tuned in an appropriate way.
Red labels filled states and blue labels empty states. The flat bands on the far left indicate the Landau levels indexed by $n$.
\label{levels}
}
\end{figure}

The scaling of the compressibility and conductivity follow from current-current correlation functions,
which do not acquire any anomalous dimensions, and thus are similar to other two-dimensional
transitions with $z = 1$ \cite{FG9087,W9255}:
\begin{align}
\Pi_{\mu\nu}(k) = (\delta_{\mu \nu} + \frac{k_\mu k_\mu}{k^2}) \Pi_e(k) + \epsilon^{\mu \nu \lambda} k_\lambda \Pi_o(k).
\end{align}
Near the critical point, $\Pi_{\mu \nu} \sim \int d^d x dt \langle J_\mu(x,t) J_\nu(0,0) \rangle \sim \xi^{1-d}$.
Therefore, $\Pi_e(k) \sim k$ and $\Pi_o(k) \sim O(k^0)$ at the critical point. From this we conclude that 
the compressibility vanishes at the critical point at zero temperature, while the conductivities are 
universal constants that can be computed in the large $N_f$ limit \cite{CF9349}.
Therefore the DC longitudinal resistivity $\rho_{xx}$ is zero on either side of the transition, but is a universal
non-zero number of order $h/e^2$ at the transition, while the DC Hall resistivity $\rho_{xy}$ is
zero on the superfluid side, $h/2e^2$ on the FQH side, and a universal number of order $h/e^2$ at the critical point. 

The temperature dependence of the polarization tensor at the critical point can
be found by replacing $k, \omega$ by $T$. It follows that the compressibility at the transition
scales like $\kappa \sim T$, while the conductivity is temperature independent at the transition. 
Finally, from general scaling considerations we can conclude that the specific heat scales like $C_v \sim T^2$. 

\it Physical realizations \rm -- The transition described here is generic, and therefore can occur in principle
in many different physical realizations involving strongly interacting systems of bosons. A particularly 
promising venue to realize bosonic FQH states is in optical traps of ultracold atoms \cite{ucaProposals},
where strongly interacting bosons in a background effective magnetic field can be realized. 
Now consider adding an external periodic potential $V_{pp}(r)$ with flux $2\pi p/q$ per plaquette. 
This induces a term $\delta H_{pp} = V_{pp}(r) b^\dagger(r) b(r)$ in the Hamiltonian of the bosons. Assuming that
the composite fermion effective theory is the correct low energy description,\footnote{This may not be the case if the gap of the parton
$f_1$ and $f_2$ bands are nearly equal.} the boson is represented by 
$b = \hat{M} f_2$, and therefore $b^\dagger b \propto f_2^\dagger f_2$, because 
$\hat{M}^\dagger \hat{M} \propto 1 +\alpha f^2 + \cdots$, where $f^2$ is the Maxwell term for $a$, $\alpha$ is a constant,
and $\cdots$ indicate higher order derivatives of the gauge field. Therefore, to leading order, the composite fermion effective
action obtains a contribution $\delta \mathcal{L}_{pp} \propto V_{pp}(r) f_2^\dagger f_2(r)$. 
Such a periodic potential may be used to induce the Chern number of the composite fermions to change. 
For small $V_{pp}$, the Landau levels split into $p$ subbands. As $V_{pp}$ is increased, the top subband
of the filled LL may eventually touch the bottom subband of the next empty LL, causing a change in the
total Chern number of the filled bands. Spatial symmetries
can force the Chern number to 
change by two units, causing a continuous FQH to superfluid transition. The necessary spatial symmetry
depends on the nature of the $V_{pp}$. There can be many ways this can happen, and the most optimal
one depends on the given experimental setup. One example is to turn on a honeycomb lattice with $2\pi$ flux per plaquette.
In the limit of large $V_{pp}$, we can pass to the tight-binding limit with nearest and next-nearest neighboring hopping,
with two low-lying bands with Chern number $\pm 1$ for the two bands \cite{H8815}. If the Chern number of the 
bottom band is $1$, it is possible in principle to achieve this regime without closing the 
energy gap. As the second neighbor hopping is tuned through zero,
there will be two band touchings, causing the Chern number to change directly from $1$ to $-1$. 
It is the $C_{3v}$ symmetry of the honeycomb lattice that protects the two Dirac cones in this case
when the second neighbor hopping is zero \cite{H8815}. 

Note that the same theory presented here can be used to develop a theory of a continuous transition between a 
chiral spin liquid and an $XY$ antiferromagnet. Also, note that the transitions considered here generically require time-reversal
symmetry to be broken, either explicitly or spontaneously, in the superfluid state before the critical point is reached. 
Finally, as with the conventional 3D XY transition, quenched disorder may significantly modify the zero temperature
phase diagram.

\it Acknowledgements \rm -- We thank Cenke Xu, T. Senthil, S. Kivelson, X.L. Qi, M. Metlitski, and M.P.A. Fisher for helpful
discussions. We especially thank Cenke Xu for referring us to \cite{RV0821} and T. Senthil, S. Kivelson for a critical reading of the
manuscript. We also acknowledge the KITP programs 
``Topological Insulators and Superconductors,'' and ``Holographic Duality and Condensed Matter Physics''
for hospitality while part of this work was done. This work was supported by a Simons Fellowship (MB)
and by the U.S. Department of Energy
(D.O.E.) under cooperative research agreement DE-FG0205ER41360,
and by the Alfred P. Sloan Foundation (JM).


\newpage
\appendix

\begin{widetext}
\section{Supplementary Material: Compressibility of slave particle/composite fermion construction of superfluid state}

In this section we will study in some more detail how the parton construction of the superfluid state 
manages to be compressible. 
As discussed in the main text, the parton construction of the superfluid state is as follows. We rewrite the boson operator as 
\begin{align}
b(r) = f_1(r) f_2(r),
\end{align}
where $f_i$ are fermions. Next, we consider a mean-field ansatz where $f_1$ forms a band insulator with Chern number $1$,
while $f_2$ forms a band insulator with Chern number $-1$, and suppose that these band insulators are created by the application
of an external periodic potential. As discussed in the main text, such a construction yields a superfluid
state for the bosons, because the emergent $U(1)$ gauge field $a$ is gapless and can be associated with the dual of the superfluid Goldstone mode. 

Since the fermions form band insulators due to an external periodic potential, by themselves they have a preferred density, 
which is set by the number of fermions per unit cell. Therefore it is not clear that the resulting state will be compressible, 
as changing the density would appear to cost a finite amount of energy. However, as we will explain below, 
such a construction does indeed yield a compressible state. 

The compressibility $\kappa$ of a zero-temperature quantum system is defined as
\begin{align}
\kappa^{-1} = \frac{\partial \mu}{\partial n} = V \frac{\partial \mu}{\partial N} = V \frac{ \partial^2 E}{\partial N^2} \sim N \frac{E(N+\delta) - 2E(N) + E(N - \delta)}{\delta^2},
\end{align}
where $\mu$ is the chemical potential, $n$ is the density, $V$ is the volume, and $E(N)$ is the ground state energy for $N$ particles,
and the above derivatives are taken at constant volume. Thus we estimate the compressibility as
\begin{align}
\kappa^{-1} \sim \frac{N \Delta_2(N,\delta)}{\delta^2},
\end{align}
where 
\begin{align}
\Delta_2(N,\delta) \sim E(N+\delta) - 2E(N) + E(N - \delta).
\end{align}
The system is incompressible if, when we take $\delta \sim \sqrt{N}$ and $N \rightarrow \infty$, $\frac{N \Delta_2(N,\delta)}{\delta^2} \rightarrow \infty$ at fixed number density.
In other words, the system is compressible if 
\begin{align}
\lim_{N \rightarrow \infty} \Delta_2(N, \sqrt{N}) < \infty.
\end{align}
The choice $\delta \sim \sqrt{N}$ is for convenience; more generally, one must take the limit $\delta, N \rightarrow \infty$ with $\delta/N \rightarrow 0$. 

In our slave-particle construction above, it was argued in the main text that the gauge field fluctuations of $a$ are gapless. 
Therefore consider a system of fermions with a filled band with a non-zero Chern number, and subject it to a magnetic field
that can vary with essentially zero energy cost. We now consider the energy $E_f(N,\phi)$, which is the ground state energy of 
the fermionic sector of the parton theory, with $N$ particles, and with additional $\phi$ flux quanta of $a$ added to the system. 
Since the flux $\phi$ is a dynamical quantity, and the gauge field $a$ is gapless, the ground state energy 
$E(N+\delta) \approx E_f(N+\delta,\delta)$, where the optimal $\phi \sim \delta$ is approximately the additional number of particles added to the system. 

Now, we would like to know the fate of
\begin{align}
\kappa^{-1} \sim \frac{N \Delta_2^\phi(N,\delta)}{\delta^2},
\end{align}
where now
\begin{align}
\Delta_2^\phi(N,\delta) \sim E_f(N+\delta, \delta) - 2E_f(N,0) + E_f(N - \delta, -\delta).
\end{align}
When the fermions fill a Landau level, $\Delta_2^\phi(N, \sqrt{N}) < \infty$ as $N \rightarrow \infty$. This is because the ground state energy of a filled lowest Landau level
is $e B N/2m$, where we set $\hbar =  c = 1$. From this, it follows that $\Delta_2^\phi(N, \delta) \sim \frac{\delta^2}{A}$, where $A$ is the area of the system,
so that $\kappa^{-1} \sim N/A$, which is bounded as $N \rightarrow \infty$ 
at fixed average number density $N/A$. Thus the Landau level problem gives a compressible state, if we allow the 
magnetic field to vary arbitrarily. This makes sense, since the density is only tied to the magnetic field, and once the magnetic field can vary arbitrarily, 
so can the density. 

Now consider a Chern insulator, such as Haldane's honeycomb model with the lowest band filled \cite{H8815}. We would like to know whether
\begin{align}
\label{cond}
\lim_{N \rightarrow \infty} \Delta_2^\phi(N, \sqrt{N})  < \infty. 
\end{align}
If so, we can then conclude that the parton Chern insulator construction of the superfulid will also be compressible if the gauge field 
$a$ is gapless. 

To establish that (\ref{cond}) is true for such a situation, consider a continuum system with a constant magnetic field, \it ie \rm a Landau level problem,
and consider adding a small periodic potential. Let $\lambda$ parametrize the strength of the periodic potential, and consider $\Delta_2^\phi(N,\delta,\lambda)$,
where the last argument just parametrizes the value of $\lambda$ in the Hamiltonian. Clearly for small $\lambda \ll e B/m$, we must have
\begin{align}
\lim_{N \rightarrow \infty} \Delta_2^\phi(N, \sqrt{N},\lambda)  < \infty. 
\end{align}
Furthermore, as long as we do not close the energy gap, continuously changing $\lambda$ must always preserve the above inequality. This is because as long
as we do not close the energy gap, the ground state energy in the thermodynamic limit is analytic in $\lambda$, and so the above inequality must 
continue to be satisfied as $\lambda$ is changed infinitesimally. 

Now, we know that it is possible to, for instance, slowly turn on a honeycomb lattice potential with $2\pi$ flux per plaquette, such that even in the limit that the periodic
potential is much stronger than $eB/m$, we do not close the energy gap. In this limit, we end up with two bands, and if the lower band has Chern number +1, 
then it is possible to adiabatically evolve from the continuum Landau level to this situation. 
For the Chern insulator with the lower band having $C = 1$
and $2\pi$ flux per plaquette, it follows that (\ref{cond}) is satisfied, because we never had to close the energy gap as we increased the periodic potential. 
Flipping the sign of the second nearest neighbor hopping in such a model can flip the Chern number. We expect therefore that as long as $C = 1$ or $C = -1$ for
the bottom band, that (\ref{cond}) will remain true. 

We conclude that Chern insulators, in addition to filled Landau levels, will satisfy (\ref{cond}) and are therefore
compressible if the magnetic field is allowed to vary arbitrarily. Since the fluctuations of the emergent $U(1)$ gauge field
are gapless in the parton construction of the superfluid, the magnetic field can indeed vary arbitrarily, so we see that the parton construction
of the superfluid state is indeed compressible when the gauge fluctuations are taken into account.  

\vskip.2in
\end{widetext}

\end{document}